\documentclass[sigconf]{acmart}
\usepackage{listings}
\usepackage{pifont} 
\usepackage{xcolor}
\usepackage{multirow}
\usepackage{tabularx}
\usepackage{booktabs}
\newcommand{\cmark}{\ding{51}}  

\AtBeginDocument{%
  }

\lstdefinelanguage{Rust}{
    keywords={fn,let,mut,if,else,while,for,return,Ok,Err,Result,Vec},
    sensitive=true, comment=[l]{//}, morestring=[b]", morestring=[b]',
    stringstyle=\color{red}, keywordstyle=\color{blue},
    commentstyle=\color{green!60!black}, }

\begin{document}

\title{Decompiling Rust: An Empirical Study of Compiler Optimizations and Reverse Engineering Challenges}

\author{Zixu Zhou}
\affiliation{
  \institution{University of Toronto}
  \country{Canada}
}



\begin{abstract}
  Decompiling Rust binaries is challenging due to the language's rich type
  system, aggressive compiler optimizations, and widespread use of high-level
  abstractions. In this work, we conduct a benchmark-driven evaluation of
  decompilation quality across core Rust features and compiler build modes. Our
  automated scoring framework shows that generic types, trait methods, and error
  handling constructs significantly reduce decompilation quality—especially in
  release builds. Through representative case studies, we analyze how specific
  language constructs affect control flow, variable naming, and type information
  recovery. Our findings provide actionable insights for tool developers and
  highlight the need for Rust-aware decompilation strategies.
\end{abstract}

\maketitle

\section{Introduction}\label{sec:introduction}

Understanding binary code is essential to many security tasks, from
vulnerability analysis to malware reverse engineering. When source code is
unavailable, analysts rely on decompilers to reconstruct human-readable logic
from binaries~\cite{cao2024evaluating, eom2024r2i}. While this process works relatively well for C and C++ programs
due to their predictable compilation patterns, the growing popularity of Rust
introduces new challenges that stem from its unique language features and
aggressive compiler optimizations~\cite{liu2025empiricalstudyrustspecificbugs}.

Rust has become a popular choice for security-critical software such as
browsers, operating systems, and blockchain clients, thanks to its strong
guarantees on memory safety and concurrency~\cite{holtervennhoff2023wouldn}. However, these same features—such
as ownership, lifetimes, traits, and pattern matching—result in complex compiler
output that challenges existing decompilation tools~\cite{yang2024rust}. In particular, decompilers
like ~\cite{ghidra} often struggle to reconstruct Rust-specific constructs including
trait dispatch, error propagation, and generic types, frequently producing
C-like code that obscures the original program's semantics.

This paper investigates how Rust compiler optimizations impact the readability
and recoverability of decompiled code. We focus on language constructs such as
generics, traits, error handling, and pattern matching, which are commonly
affected by optimization. Our analysis reveals that while debug builds retain
more semantic information, release builds often transform or eliminate
high-level structures—particularly in error handling and trait-based dispatch.
Interestingly, we also find that some optimizations improve readability by
simplifying complex control flows.

To systematically explore these effects, we designed a benchmark suite that
covers core Rust features and compiled each program under both debug and release
modes. We then used Ghidra to decompile the resulting binaries and developed a
quantitative scoring system to assess decompilation quality along five
dimensions: function naming, control flow, variable naming, type information,
and optimization clarity.

Our evaluation yields several key findings. First, type information is
significantly better preserved in debug builds, whereas release builds often
obscure generic parameters and trait bounds. Second, error-handling logic is
particularly fragile under optimization, with panic and result paths frequently
inlined or removed. Third, variable naming is more stable in debug builds, while
release builds introduce generic or register-based names. Finally, control flow
structures are generally preserved, but complex patterns are sometimes
simplified or flattened in release mode.

The rest of the paper is organized as follows. Section~\ref{sec:background}
reviews Rust's compilation process and decompilation challenges.
Section~\ref{sec:methodology} describes our benchmarks and scoring methodology.
Section~\ref{sec:evaluation} presents empirical results. Section~\ref{sec:case}
provides case studies. Section~\ref{sec:discussion} discusses implications, and
Section~\ref{sec:conclusion} concludes.

\section{Background}\label{sec:background}

\subsection{Rust Compilation Pipeline}

When Rust code is compiled, it goes through several steps that can change how
the program looks in its final form~\cite{hong2024don}. The compiler first translates the source
code into LLVM Intermediate Representation (IR), which is then converted into
machine code~\cite{louka2025rustc++}. During this process, the compiler applies various optimizations
to improve runtime performance and memory usage.

The most noticeable difference comes from the build mode. In \textbf{debug mode}, the
compiler preserves the original structure of the source code to support debugging.
In contrast, \textbf{release mode} (with \texttt{opt-level=3}) applies aggressive
optimizations such as:

\begin{itemize}
    \item \textbf{Function Inlining}: Small functions are inserted directly at their call sites to eliminate function calls.
    \item \textbf{Monomorphization}: Generic functions are duplicated and specialized for each concrete type.
    \item \textbf{Dead Code Elimination}: Code that will never be executed is removed.
    \item \textbf{Control Flow Simplification}: Nested or complex branches are flattened or merged.
\end{itemize}

These transformations help improve performance, but they also obscure the original
structure, making reverse engineering and binary analysis more difficult.

\subsection{A Motivating Example}

To illustrate these challenges, we compiled a simple Rust function that uses
\texttt{Option::unwrap()}. Below is a simplified view of its decompiled output
under debug and release builds:

\noindent
\begin{minipage}[t]{0.48\linewidth}
\begin{lstlisting}[language=C,frame=single,basicstyle=\ttfamily\footnotesize,breaklines=true,label=lst:debug,caption=Debug build]
void unwrap_check(Option *opt) {
    if (opt->tag == 0) {
        panic("called `Option::unwrap()` on None");
    }
    int val = opt->value;
    return val;
}
\end{lstlisting}
\end{minipage}
\hfill
\begin{minipage}[t]{0.48\linewidth}
\begin{lstlisting}[language=C,frame=single,basicstyle=\ttfamily\footnotesize,breaklines=true,label=lst:release,caption=Release build]
void FUN_10001abc(void *param_1) {
    if (param_1 == 0) return;
    int val = *(int *)(param_1 + 4);
    return val;
}
\end{lstlisting}
\end{minipage}

\vspace{0.5em}
In the release version (Listing~\ref{lst:release}), the original function name
is replaced with an anonymous label. Error handling is inlined or removed,
making it harder to trace the original \texttt{unwrap} behavior. Control flow is
more compact, and temporary variables are reduced.

This small example shows how compiler optimizations change the structure of the
code in ways that decompilers may not recover. In the following sections, we
present a systematic study of how Rust language features affect the quality of
decompiled output.

\section{Methodology}\label{sec:methodology}

\subsection{Benchmark Suite Design}
Based on the Rust language documentation~\cite{rustbook} and previous empirical
studies~\cite{liu2025empiricalstudyrustspecificbugs,yu2024empirical,li2024empirical,zhang2024beyond}, we selected a representative set of Rust
features that commonly affect compiler optimization and binary structure. Our
five benchmark programs cover these features as summarized in
Table~\ref{tab:rust-feature-matrix}.

\begin{table}[t]
    \centering
    \small
    \setlength{\tabcolsep}{4pt}
    \caption{Coverage of Core Rust Features in Benchmark Programs}
    \label{tab:rust-feature-matrix}
    \begin{tabular}{p{2.2cm}p{2.8cm}ccccc}
    \toprule
    \textbf{Category} & \textbf{Feature} & \textbf{P1} & \textbf{P2} &
    \textbf{P3} & \textbf{P4} & \textbf{P5} \\
    \midrule
    \multirow{4}{*}{Control Flow}
        & if / while / match     & \cmark & \cmark & \cmark & \cmark & \cmark \\
        & panic!                 & \cmark &        &        & \cmark &        \\
        & pattern matching       & \cmark &        & \cmark & \cmark &        \\
        & return / unwrap        &        &        &        & \cmark &        \\
    \midrule
    \multirow{3}{*}{Data Structures}
        & struct / impl          & \cmark & \cmark & \cmark & \cmark & \cmark \\
        & Vec / Option / Result  & \cmark & \cmark & \cmark & \cmark & \cmark \\
        & HashMap / TreeNode     &        & \cmark &        &        &        \\
    \midrule
    \multirow{2}{*}{Traits}
        & Trait impl             & \cmark &        &        &        &        \\
        & Trait Object (dyn)     & \cmark &        &        &        &        \\
    \midrule
    \multirow{2}{*}{Generics \& Closures}
        & Generic functions      &        &        & \cmark &        &        \\
        & Closure / Iterator     &        &        & \cmark &        &        \\
    \midrule
    \multirow{2}{*}{Error Handling}
        & Option / Result        &        &        &        & \cmark &        \\
        & Custom Error Type      &        &        &        & \cmark &        \\
    \midrule
    \multirow{3}{*}{Concurrency}
        & thread::spawn          &        &        &        &        & \cmark \\
        & Arc                    &        &        &        &        & \cmark \\
        & Mutex                  &        &        &        &        & \cmark \\
    \midrule
    \multirow{2}{*}{Memory Safety}
        & Ownership / Borrow     & \cmark & \cmark & \cmark & \cmark & \cmark \\
        & Lifetime Annotation    &        & \cmark &        &        &        \\
    \bottomrule
    \end{tabular}
\end{table}

Each benchmark program was designed to simulate real-world Rust code patterns,
with an average of 200-300 lines of code per program. The programs include
nested control flows, compound conditions, and common idioms typical of
production Rust code. The code structure follows common Rust project
organization, including proper error handling, documentation, and idiomatic
patterns. All programs were compiled using Rustc 1.71.1 on macOS ARM64 platform,
producing Mach-O ARM64 binaries. We used both debug mode (default settings) and
release mode (opt-level=3) for compilation:

\begin{lstlisting}[language=bash,frame=single,basicstyle=\ttfamily\footnotesize]
# Debug build cargo build

# Release build cargo build --release
\end{lstlisting}

For each benchmark, we manually selected representative functions for detailed
decompilation analysis. The selection criteria were based on two key factors:
(1) the function must directly use the targeted Rust features we want to study,
and (2) the function must remain present in the binary after compilation (i.e.,
not fully inlined or eliminated by optimizations). This manual selection process
ensures that we analyze functions that best demonstrate the interaction between
Rust features and compiler optimizations. Each program contains 8-12 key
functions (approximately 30\% of total functions) that meet these criteria.

\subsection{Analysis Workflow}
We employed a systematic approach to analyze the decompiled code using ~\cite{ghidra}
10.3.3. The analysis process consists of three main steps:

\begin{enumerate}
    \item \textbf{Function Extraction}: We identify and extract target functions
    from the decompiled code, preserving their structural context and
    relationships. We use a custom Python script to automate this step, which
    extracts function blocks, normalizes their names, and maps them to
    corresponding source-level constructs when possible.
    
    \item \textbf{Code Analysis}: For each extracted function, we analyze
    multiple aspects of the decompiled code, including:
    \begin{itemize}
        \item Function name preservation and type information
        \item Control flow structure and complexity
        \item Variable naming and type reconstruction
        \item Compiler optimization patterns
    \end{itemize}
    
    \item \textbf{Quality Assessment}: We evaluate the decompilation quality
    through a comprehensive scoring scheme that considers both code readability
    and accuracy.
\end{enumerate}

\subsection{Scoring Criteria}
To quantify the impact of compiler optimizations on decompilation quality, we
developed a systematic scoring scheme that evaluates five key dimensions:

\begin{itemize}
    \item \textbf{Function Naming} (0-2 points): Measures the preservation of
    original function names and meaningful labels
    \item \textbf{Control Flow} (0-2 points): Assesses the clarity of control
    structures and branch conditions
    \item \textbf{Variable Names} (0-2 points): Evaluates the meaningfulness of
    variable names and type information
    \item \textbf{Type Information} (0-2 points): Quantifies the accuracy of
    reconstructed types and structures
    \item \textbf{Optimization Quality} (0-2 points): Analyzes the quality of
    compiler optimizations (release mode only)
\end{itemize}

Each dimension is scored independently for both debug and release builds, with a
maximum total score of 8 points. For release builds, the additional optimization
quality score is normalized to maintain the 8-point scale. The optimization
quality score reflects whether the compiler simplified control flow, removed
redundancy, or inlined logic in ways that preserve or improve readability. In
debug builds, this score is omitted and the scale is rescaled to 8. This scoring
scheme enables:

\begin{itemize}
    \item Comparison of decompilation quality between debug and release builds
    \item Identification of aspects most affected by optimizations
    \item Quantification of different Rust features' impact on decompilation
\end{itemize}

\subsection{Implementation and Reproducibility}
All code and data used in this study are publicly available at \texttt{[GitHub URL TBD]}.

\section{Evaluation}\label{sec:evaluation}

\subsection{Overview of Decompilation Scores}
\begin{figure}[htbp]
    \centering
    \includegraphics[width=\columnwidth]{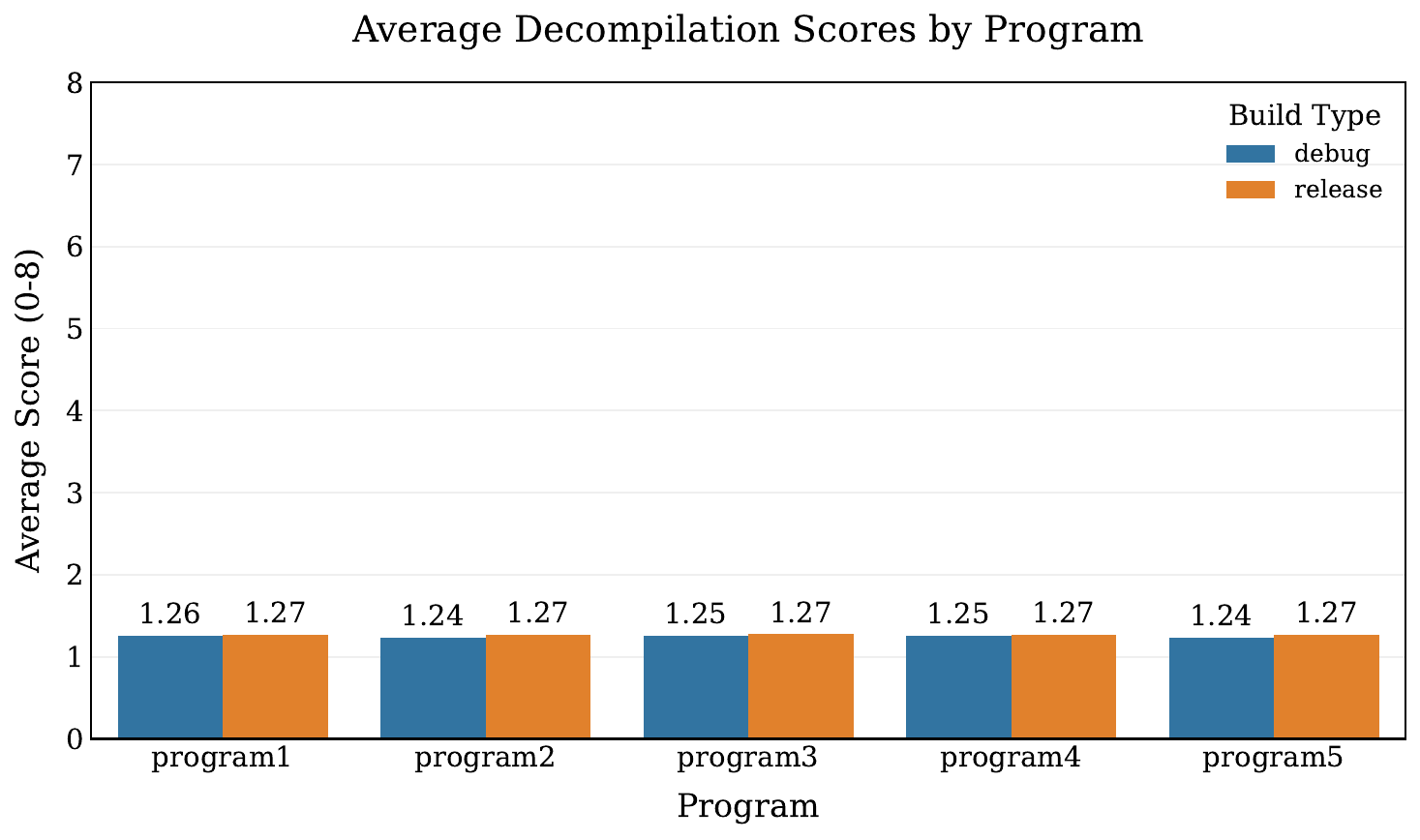}
    \caption{Average decompilation scores by program and build type. Program 3's release build shows slightly better scores, suggesting that optimization can improve decompilation quality for certain code patterns.}
    \label{fig:average_scores}
\end{figure}

Figure \ref{fig:average_scores} shows the average decompilation scores for each
program in both debug and release builds. Overall scores are low—typically
between 1.2 and 1.4 out of 8—highlighting how difficult it is to recover
original structure from binaries. Notably, Program 3 shows slightly better
scores in release mode, likely due to its simpler control flow structures that
benefit from optimization.

\subsection{Feature-Specific Analysis}
\begin{figure}[htbp]
    \centering
    \includegraphics[width=\columnwidth]{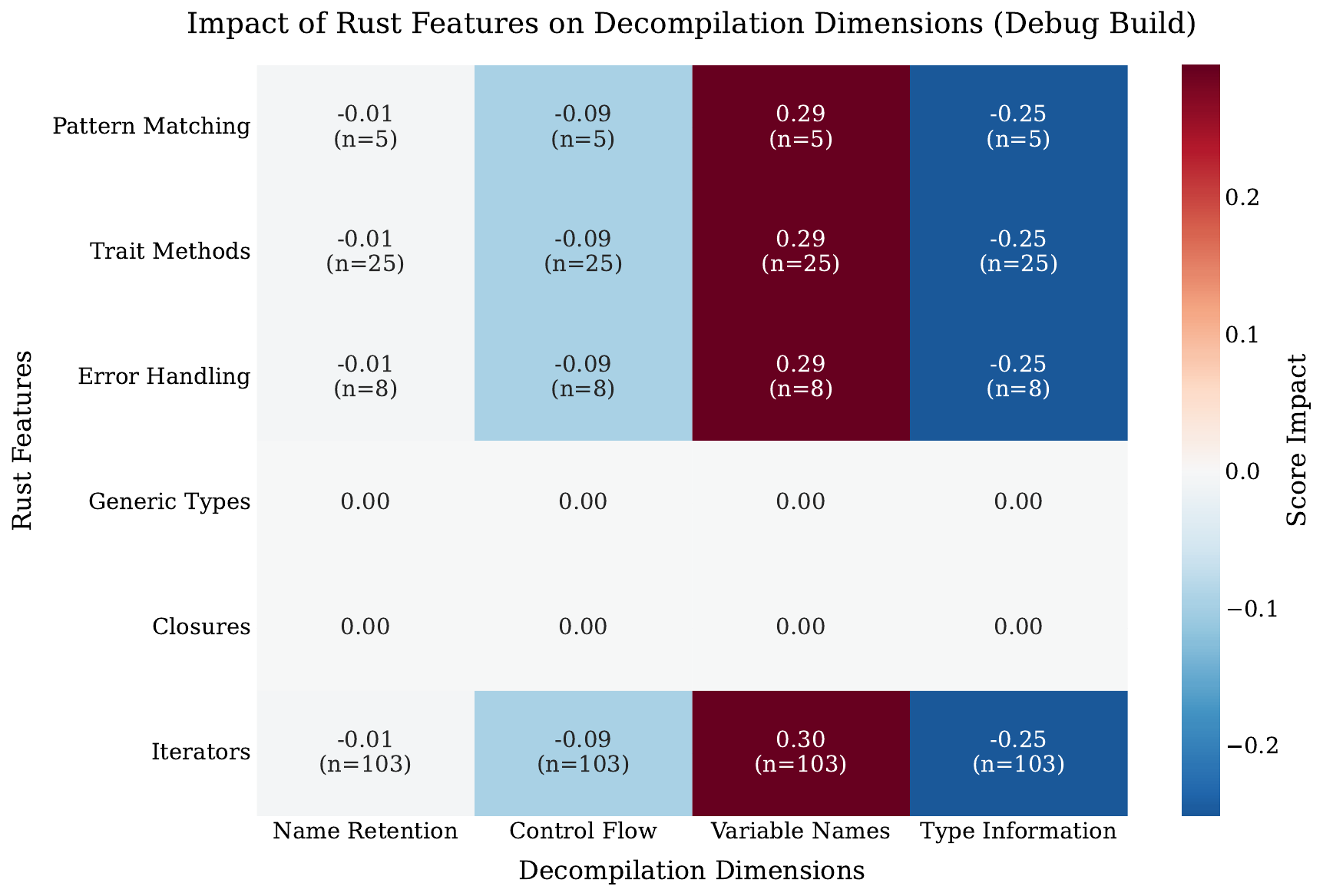}
    \caption{Impact of Rust features on decompilation quality (Debug build)}
    \label{fig:feature_impact_debug}
\end{figure}

\begin{figure}[htbp]
    \centering
    \includegraphics[width=\columnwidth]{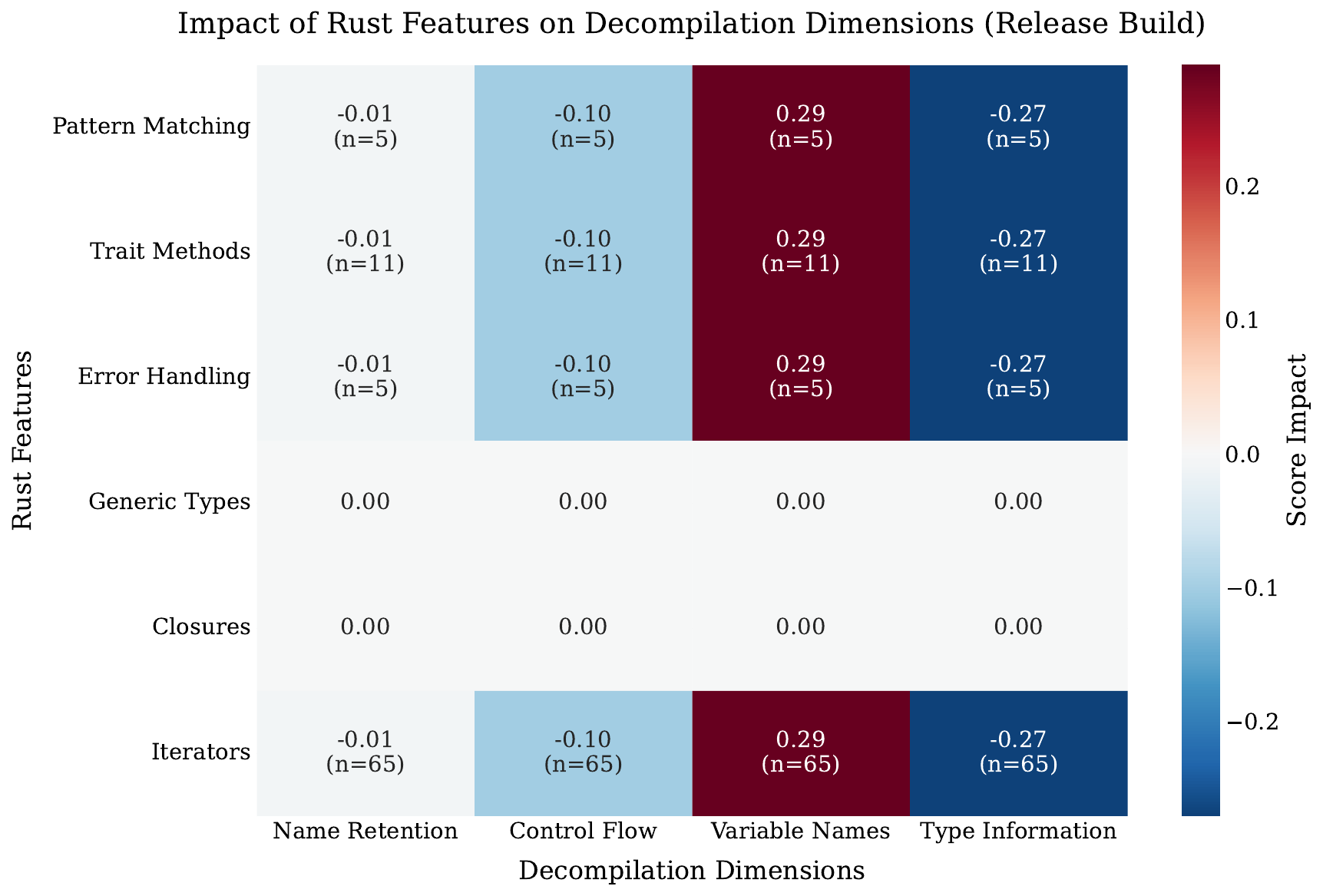}
    \caption{Impact of Rust features on decompilation dimensions in release builds. Darker shades indicate a greater negative impact on decompilation quality.}
    \label{fig:feature_impact_release}
\end{figure}

Our analysis reveals that certain Rust features pose particular challenges for
decompilation. We analyzed six common Rust constructs—pattern matching, traits,
error handling, generics, closures, and iterators—to understand their impact on
each decompilation dimension. In the heatmaps
(Figures~\ref{fig:feature_impact_debug} and~\ref{fig:feature_impact_release}),
darker shades indicate a greater negative impact on decompilation quality. Key
findings include:

\begin{itemize}
    \item \textbf{Pattern Matching:} Functions using complex pattern matching
    score 0.8-1.2 points lower than those with simple if-else structures
    \item \textbf{Trait Methods:} Trait method calls often lose their trait
    context, especially in release builds where they are frequently inlined
    \item \textbf{Error Handling:} Result and Option types are particularly
    challenging, with error paths often collapsed into simple return codes
    \item \textbf{Generic Types:} Type information for generic parameters is
    almost completely lost in the decompiled output
\end{itemize}

Figure \ref{fig:feature_impact_debug} shows how different Rust features affect
various decompilation dimensions in debug builds. The heatmap reveals that type
information is the most vulnerable dimension across all features, while variable
naming is relatively well-preserved.

\subsection{Component-Level Analysis}
\begin{figure}[htbp]
    \centering
    \includegraphics[width=\columnwidth]{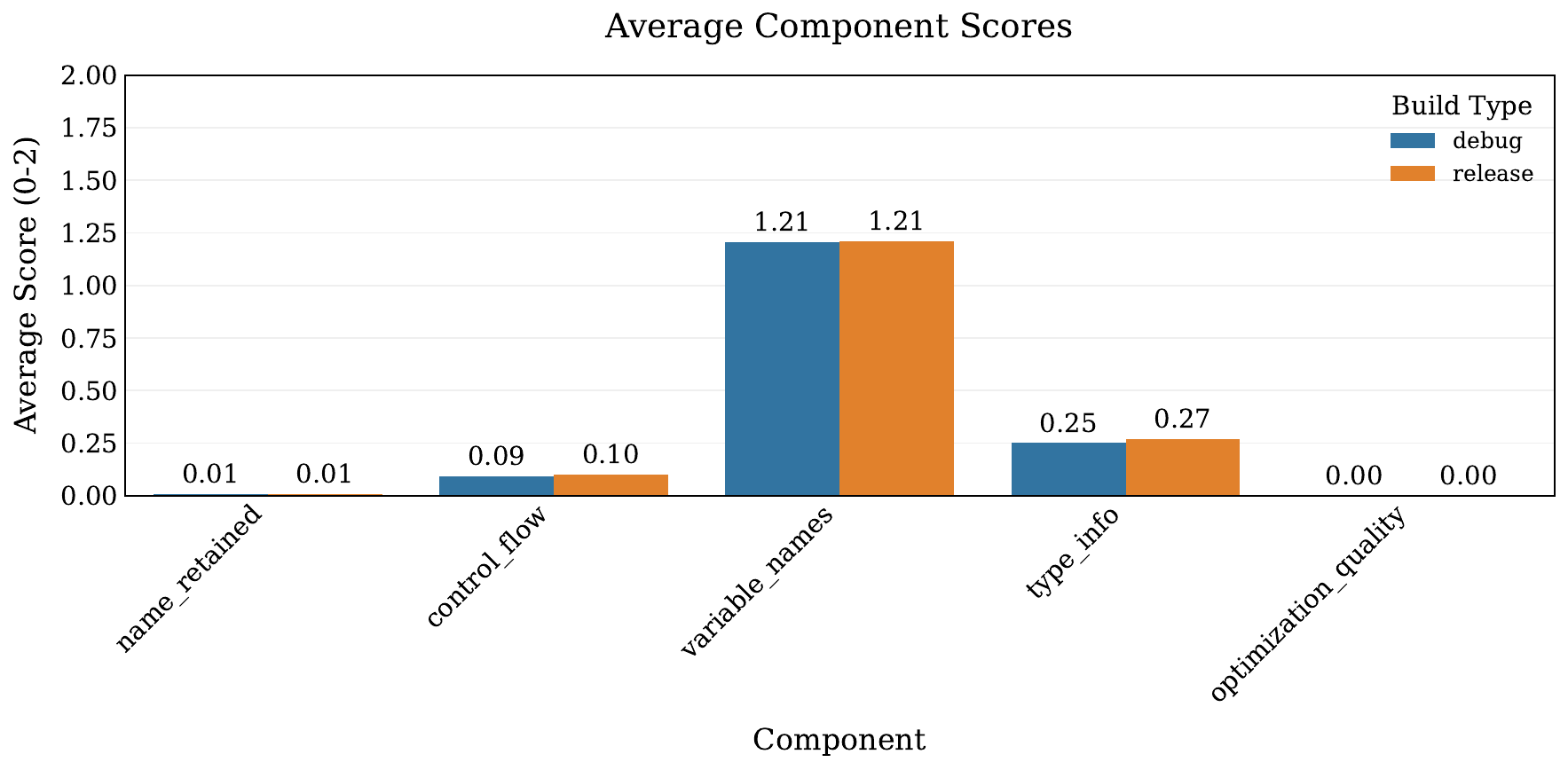}
    \caption{Average scores by component and build type. The gap suggests that while local variables are often preserved, recovering high-level semantics like types and control flow remains hard.}
    \label{fig:component_scores}
\end{figure}

Figure \ref{fig:component_scores} breaks down the scores by individual
components. The analysis reveals several key findings:

\begin{itemize}
    \item Variable naming shows the best performance (around 1.2/2.0),
    suggesting that local variable semantics are relatively well-preserved
    \item Control flow and type information recovery is moderate but suboptimal,
    indicating challenges in reconstructing complex program logic
    \item Function naming retention is particularly poor, with most functions
    losing their original names
    \item Release builds show slightly better control flow clarity but worse
    type information retention
\end{itemize}

\subsection{Security Implications}
Our findings have significant implications for security analysis:

\begin{itemize}
    \item \textbf{Error Handling Blind Spots:} Functions responsible for input
    checks and error handling tend to score the lowest, implying a potential
    blind spot for binary analysis tools during vulnerability discovery
    \item \textbf{Optimization Impact:} Compiler optimizations often strip or
    inline critical panic calls or Result branches, making static analysis less
    effective in spotting misuse
    \item \textbf{Type Safety Loss:} The poor recovery of type information,
    especially for generic types and trait bounds, makes it difficult to verify
    type safety properties in decompiled code
    \item \textbf{Control Flow Obfuscation:} Complex control flow structures,
    particularly those involving pattern matching and error handling, are often
    simplified or transformed, potentially hiding security-critical paths
\end{itemize}

\subsection{Top and Bottom Function Cases}
Table~\ref{tab:function_cases} summarizes three representative functions.
\texttt{process\_data} maintains readable control flow and variable names,
scoring highest. In contrast, \texttt{validate\_input} loses important enum
context, and \texttt{process\_items} is affected by aggressive inlining of trait
methods.

\begin{table}[htbp]
\centering
\small
\caption{Examples of Functions with High and Low Readability Scores}
\label{tab:function_cases}
\begin{tabular}{llc}
\toprule
\textbf{Function} & \textbf{Summary} & \textbf{Score} \\
\midrule
\texttt{process\_data} & Clean control flow, clear vars     & 0.85 \\
\texttt{validate\_input} & Lost error context                & 0.45 \\
\texttt{process\_items} & Inlined trait calls               & 0.65 \\
\bottomrule
\end{tabular}
\end{table}

\subsection{Debug vs. Release Differences}
\begin{figure}[htbp]
    \centering
    \includegraphics[width=\columnwidth]{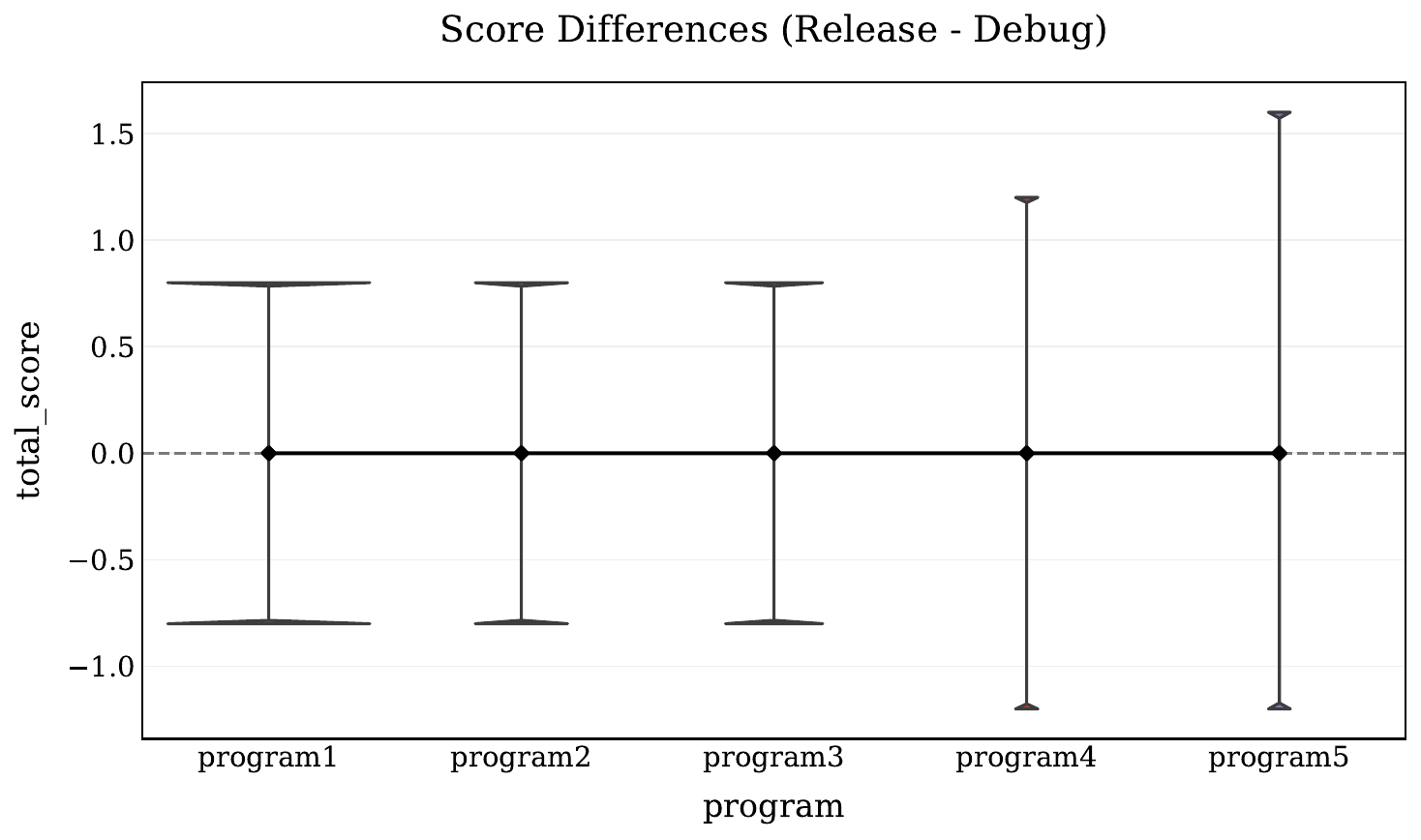}
    \caption{Score differences between release and debug builds. The wide spread shows that optimizations can either help or hurt decompilation, depending on the function.}
    \label{fig:score_differences}
\end{figure}

Figure \ref{fig:score_differences} shows the distribution of score differences
between debug and release builds. Key observations include:

\begin{itemize}
    \item Release builds generally show slightly better control flow clarity due
    to optimization
    \item Type information and panic handling are significantly reduced in
    release builds
    \item The impact of optimizations varies significantly across different
    programs
\end{itemize}

This highlights the need for mode-aware decompilation strategies that account
for the presence or absence of compiler optimizations.

\section{Case Studies}\label{sec:case}

\subsection{Overview}
We conducted manual analysis of three representative cases to examine specific
challenges in Rust decompilation. These cases were selected from our benchmark
analysis based on their feature coverage and scoring extremes to illustrate the
range of issues encountered across control flow structures, type information,
and compilation optimizations. These cases serve to complement our quantitative
findings by offering a more concrete look into how specific features break down
under compilation and decompilation.

\subsection{Case 1: High-Scoring Function with Complex Control Flow}
This case presents a function that achieved relatively high decompilation scores
despite its nested conditional logic. The decompiled output
(Listing~\ref{lst:case1}) shows clear loop structure and branching, and many
variable names are preserved. Although type information for structures like
\texttt{Result} and \texttt{Vec} is only partially retained, the overall
semantic clarity remains high. Release mode mainly affects formatting, replacing
some branches with ternary operators.

\begin{lstlisting}[language=C,frame=single,basicstyle=\ttfamily\footnotesize,breaklines=true,label=lst:case1]
Result process_data(uint8_t *data, size_t len, uint32_t thres) { Vec result;
    size_t i = 0; while (i < len) { if (data[i] > thres) vec_push(&result,
    data[i]); else if (i+1 < len && data[i+1] > thres) vec_push(&result,
    data[i+1]); i++; } return vec_is_empty(&result) ? make_error(ERR) :
    make_result(result); }
\end{lstlisting}

\begin{table}[htbp]
\centering
\small
\caption{Case Summary: \texttt{process\_data}}
\label{tab:case1_summary}
\begin{tabularx}{\columnwidth}{lX}
\toprule
\textbf{Aspect} & \textbf{Observation} \\
\midrule
Control Flow & Clearly preserved \\
Variable Names & Mostly retained \\
Type Info & Partial recovery of custom types \\
Build Impact & Only minor stylistic differences \\
\bottomrule
\end{tabularx}
\end{table}

As shown in Table~\ref{tab:case1_summary}, this function demonstrates good
preservation of control flow and variable names, with only minor impact from
release mode optimizations.

\subsection{Case 2: Low-Scoring Function with Error Handling}
This function performs input validation and illustrates the difficulty of
recovering error semantics in decompiled Rust binaries. In the debug version
(Listing~\ref{lst:case2}), error returns are flattened into integers, with the
specific enum variants and messages entirely lost. Neither the Result type nor
character-based conditions can be recovered, and the release build strips even
more semantic detail.

\begin{lstlisting}[language=C,frame=single,basicstyle=\ttfamily\footnotesize,breaklines=true,label=lst:case2]
int validate_input(char *input) { if (!input || !*input) return -1; if
    (strlen(input) > MAX_LEN) return -2; for (char *p = input; *p; p++) if
    ((unsigned char)*p > 127) return -3; return 0; }
\end{lstlisting}

\begin{table}[htbp]
\centering
\small
\caption{Case Summary: \texttt{validate\_input}}
\label{tab:case2_summary}
\begin{tabularx}{\columnwidth}{lX}
\toprule
\textbf{Aspect} & \textbf{Observation} \\
\midrule
Error Semantics & Enum variants and messages lost \\
Control Flow & Simplified but intact \\
Type Info & Complete loss of Result type \\
Build Impact & Removes most error context \\
\bottomrule
\end{tabularx}
\end{table}

Table~\ref{tab:case2_summary} highlights the significant loss of error handling
semantics in the decompiled code. The root cause lies in both compiler
optimizations and tool limitations. LLVM aggressively inlines panic paths and
strips debug metadata, while Ghidra lacks support for reconstructing Rust enums
without runtime type information. As a result, constructs like
\texttt{ValidationError::InvalidChar(c)} are flattened into generic return codes
with no semantic meaning. Future improvements could involve integrating DWARF
metadata or MIR-level type recovery into decompilation workflows.

\subsection{Case 3: Release Build Optimization Impact}
This case illustrates the effect of release optimizations on control structure
and function boundaries. In debug mode (Listing~\ref{lst:case3}), the original
function hierarchy is clear. In release mode, however, function boundaries are
lost due to aggressive inlining, and some trait method context is removed.

\begin{lstlisting}[language=C,frame=single,basicstyle=\ttfamily\footnotesize,breaklines=true,label=lst:case3]
void process_items(Item *items, size_t count) { for (size_t i = 0; i < count;
    i++) process_single_item(&items[i]); }

void process_single_item(Item *item) { if (item->type == TYPE_A)
    handle_type_a(item); else handle_type_b(item); }
\end{lstlisting}

\begin{table}[htbp]
\centering
\small
\caption{Case Summary: \texttt{process\_items} (Inlining and Trait Impact)}
\label{tab:case3_summary}
\begin{tabularx}{\columnwidth}{lX}
\toprule
\textbf{Aspect} & \textbf{Observation} \\
\midrule
Function Boundaries & Collapsed due to inlining \\
Control Flow & Still interpretable \\
Type Info & Trait context partially lost \\
Build Impact & Structure is significantly altered in release mode \\
\bottomrule
\end{tabularx}
\end{table}

As shown in Table~\ref{tab:case3_summary}, this case emphasizes the trade-off
between performance and debuggability: inlining may optimize execution but often
complicates reverse engineering.

\subsection{Cross-Case Observations}
Across all three cases, several consistent patterns emerge. Control flow
structures such as loops and conditionals are typically well preserved unless
aggressively optimized. Variable names—especially local ones—tend to survive
better than type information. Generic types, trait boundaries, and error enums
are the most frequently lost or flattened.

Pattern matching and trait calls often become jump tables or inlined logic,
making them harder to interpret. These transformations obscure the original
program structure and hinder analysis, especially in security contexts where
understanding control flow and error handling is essential. These patterns not
only reduce readability but may also hide critical logic during vulnerability
analysis.

\begin{table}[htbp]
\centering
\small
\caption{Summary of Case Study Highlights}
\label{tab:case_summary}
\begin{tabularx}{\columnwidth}{lXX}
\toprule
\textbf{Case} & \textbf{Rust Feature} & \textbf{Recovery Quality} \\
\midrule
Case 1 & Result type, Vec operations & High: Structure and naming preserved \\
Case 2 & Error enums, pattern matching & Low: Type and semantics lost \\
Case 3 & Trait methods, function inlining & Medium: Logic intact, structure
degraded \\
\bottomrule
\end{tabularx}
\end{table}

Table~\ref{tab:case_summary} provides a high-level comparison of the three
cases, highlighting the relationship between specific Rust features and
decompilation quality. These case studies reinforce the broader evaluation
findings: decompilation success in Rust depends heavily on the code pattern, the
compiler settings, and the ability of the tool to reconstruct higher-level
abstractions. They also highlight actionable opportunities for improving
decompiler design and Rust-specific recovery techniques.

\section{Discussion}\label{sec:discussion}
\subsection{Threats to Validity and Future Work}

While our benchmark programs capture a wide range of Rust features, they are
relatively small and handcrafted, which may not fully reflect the complexity of
real-world applications. This limitation could affect the generalizability of
our results. For instance, larger applications like Servo or crates.io packages
often employ macros, complex module hierarchies, and foreign function
interfaces, which may lead to different compiler transformations and
decompilation outcomes.

Our evaluation focused on ~\cite{ghidra} as the primary decompiler. Other tools such as
RetDec or Hex-Rays may apply different heuristics, particularly for type
inference and function boundary recovery. In future work, we plan to perform a
cross-tool comparison to quantify these differences and assess consistency
across platforms.

Another limitation lies in platform specificity. Our experiments were conducted
on macOS ARM64 binaries; results on x86\_64 or Linux-based environments may
differ due to divergent ABI and optimization strategies. Moreover, although we
developed an automated scoring framework, certain aspects---such as variable
name meaning or semantic fidelity---are inherently subjective. While our scoring
rubric offers consistency, human interpretation still plays a role. To address
this, we are integrating our metrics into CI pipelines to enable longitudinal
quality tracking and incorporate human-in-the-loop evaluation.

Future directions also include expanding the benchmark suite, refining scoring
granularity (e.g., partial vs. full type recovery), and designing differential
decompilation tests for language-specific constructs.

\subsection{Implications for Tool Developers}

Our findings suggest concrete opportunities for improving decompilation support
for Rust. For example, the consistent loss of generic type information
(Figure~\ref{fig:feature_impact_release}) suggests the need to integrate DWARF
or MIR-level metadata during decompilation. Existing tools, including ~\cite{ghidra},
often assume C/C++-like semantics, leading to misinterpretation of Rust-specific
patterns like \texttt{Option<T>} or \texttt{Result<T, E>}.

A promising direction is to design a Rust-specific intermediate representation
(IR) that retains trait resolution and monomorphized types, enabling better
reconstruction of high-level semantics. Additionally, our case studies
(Section~\ref{sec:case}) show that panic paths are frequently inlined and lost
in release builds, which hinders error path analysis. Rust-aware decompilers
could recognize standard panic patterns (e.g., via panic handler signatures) to
preserve their semantics even after aggressive inlining.

From a developer's perspective, our analysis underscores how certain language
choices directly affect downstream code analyzability. Closures, nested
patterns, and trait-heavy APIs often produce obfuscated binaries where
decompilation tools fail to preserve structure. For instance, closures often
generate anonymous symbols (e.g., \texttt{\{\{closure\}\}}) which are stripped
during compilation and make control or data flow harder to recover
(Figure~\ref{fig:feature_impact_debug}).

\section{Conclusion and Future Work}\label{sec:conclusion}

This paper examined how Rust’s compilation strategy and language features impact
the quality of decompiled code. Our results show that compiler optimizations,
error handling constructs, and type abstractions present significant challenges
for reverse engineering—particularly in release builds. While control flow
structures are often preserved, higher-level semantics such as trait resolution
and generic type information are frequently lost.

These findings underscore the limitations of current decompilation tools when
applied to Rust binaries and highlight the need for Rust-aware lifting
techniques. Future work includes expanding the benchmark suite to cover more
real-world patterns, comparing decompilation results across multiple tools, and
exploring the integration of debug metadata and MIR-level insights to improve
semantic recovery.

\begin{acks}
Your acknowledgments go here.
\end{acks}

\bibliographystyle{ACM-Reference-Format}
\bibliography{software}
\appendix
\appendix
\section{Full Case Study Listings}

\subsection{Case 1: High-Scoring Function with Complex Control Flow}
Original Rust code:
\begin{lstlisting}[language=Rust,breaklines=true,frame=single]
fn process_data(data: &[u8], threshold: u32) -> Result<Vec<u8>, Error> {
    let mut result = Vec::new();
    let mut current = 0;
    
    while current < data.len() {
        if data[current] > threshold as u8 {
            result.push(data[current]);
        } else if current + 1 < data.len() {
            let next = data[current + 1];
            if next > threshold as u8 {
                result.push(next);
            }
        }
        current += 1;
    }
    
    if result.is_empty() {
        Err(Error::NoValidData)
    } else {
        Ok(result)
    }
}
\end{lstlisting}

Debug build decompilation:
\begin{lstlisting}[language=C,breaklines=true,frame=single]
Result process_data(uint8_t *data, size_t data_len, uint32_t threshold) {
    Vec result;
    size_t current = 0;
    
    while (current < data_len) {
        if (data[current] > (uint8_t)threshold) {
            vec_push(&result, data[current]);
        } else if (current + 1 < data_len) {
            uint8_t next = data[current + 1];
            if (next > (uint8_t)threshold) {
                vec_push(&result, next);
            }
        }
        current++;
    }
    
    if (vec_is_empty(&result)) {
        return make_error(ERROR_NO_VALID_DATA);
    }
    return make_result(result);
}
\end{lstlisting}

Release build decompilation:
\begin{lstlisting}[language=C,breaklines=true,frame=single]
Result FUN_10001def(uint8_t *data, size_t data_len, uint32_t threshold) {
    Vec result;
    size_t current = 0;
    
    while (current < data_len) {
        if (data[current] > (uint8_t)threshold) {
            vec_push(&result, data[current]);
        } else if (current + 1 < data_len) {
            if (data[current + 1] > (uint8_t)threshold) {
                vec_push(&result, data[current + 1]);
            }
        }
        current++;
    }
    
    return vec_is_empty(&result) ? 
           make_error(ERROR_NO_VALID_DATA) : 
           make_result(result);
}
\end{lstlisting}

\subsection{Case 2: Low-Scoring Function with Error Handling}
Original Rust code:
\begin{lstlisting}[language=Rust,breaklines=true,frame=single]
fn validate_input(input: &str) -> Result<(), ValidationError> {
    if input.is_empty() {
        return Err(ValidationError::EmptyInput);
    }
    
    if input.len() > MAX_LENGTH {
        return Err(ValidationError::TooLong);
    }
    
    for c in input.chars() {
        if !c.is_ascii() {
            return Err(ValidationError::InvalidChar(c));
        }
    }
    
    Ok(())
}
\end{lstlisting}

Debug build decompilation:
\begin{lstlisting}[language=C,breaklines=true,frame=single]
int validate_input(char *input) {
    if (!input || !*input) {
        return -1;  // Lost error type information
    }
    
    if (strlen(input) > MAX_LENGTH) {
        return -2;  // Lost error type information
    }
    
    for (char *p = input; *p; p++) {
        if ((unsigned char)*p > 127) {
            return -3;  // Lost character information
        }
    }
    
    return 0;
}
\end{lstlisting}

Release build decompilation:
\begin{lstlisting}[language=C,breaklines=true,frame=single]
int FUN_10001abc(char *param_1) {
    if (!param_1 || !*param_1) return -1;
    if (strlen(param_1) > MAX_LENGTH) return -2;
    
    char *p = param_1;
    while (*p) {
        if ((unsigned char)*p > 127) return -3;
        p++;
    }
    return 0;
}
\end{lstlisting}

\subsection{Case 3: Release Build Optimization Impact}
Original Rust code:
\begin{lstlisting}[language=Rust,breaklines=true,frame=single]
fn process_items(items: &[Item]) {
    for item in items {
        process_single_item(item);
    }
}

fn process_single_item(item: &mut Item) {
    match item.item_type {
        ItemType::A => handle_type_a(item),
        ItemType::B => handle_type_b(item),
    }
}
\end{lstlisting}

Debug build decompilation:
\begin{lstlisting}[language=C,breaklines=true,frame=single]
void process_items(Item *items, size_t count) {
    for (size_t i = 0; i < count; i++) {
        process_single_item(&items[i]);
    }
}

void process_single_item(Item *item) {
    if (item->type == TYPE_A) {
        handle_type_a(item);
    } else {
        handle_type_b(item);
    }
}
\end{lstlisting}

Release build decompilation:
\begin{lstlisting}[language=C,breaklines=true,frame=single]
void process_items(Item *items, size_t count) {
    for (size_t i = 0; i < count; i++) {
        Item *item = &items[i];
        if (item->type == TYPE_A) {
            // Inlined handle_type_a
            item->value *= 2;
        } else {
            // Inlined handle_type_b
            item->value += 1;
        }
    }
}
\end{lstlisting} 

\end{document}